Direct Measurement of the Destruction of Charge Quantization in a Single Electron Box

David S. Duncan, Carol Livermore, and Robert M. Westervelt

Division of Applied Sciences and Department of Physics, Harvard University,

Cambridge, MA 02138, USA

Kevin D. Maranowski and Arthur C. Gossard

Materials Department, University of California, Santa Barbara

Santa Barbara, CA 93106, USA

Abstract

We report here direct measurements of the destruction of charge quantization in a single electron box, the first over the full range of box-to-lead conductance values from $G \sim 0$ to the conductance quantum $G_Q = 2e^2/h$, using a sensitive single-electron transistor (SET) electrometer. The sensitivity of the electrometer is measured to be $dq \sim 6 \times 10^{-5}$ e/$\sqrt{Hz}$ and its superiority to conductance measurements of charge fluctuations is clearly demonstrated. As the rate of quantum mechanical tunneling from the box to its lead is increased, the quantization of charge is destroyed, disappearing entirely at $G = G_Q$ in agreement with theory.

The quantization of charge plays a fundamental role in the behavior of quantum dots, which are small, isolated metallic regions connected to a circuit by tunnel junctions[1-3]. Understanding the conditions necessary for charge quantization in such a system is an important topic in many-body theory, analogous to the Kondo problem for spin[4-5]. The consequences of charge quantization and related mesoscopic phenomena are also of increasing concern to the microelectronics industry[6]. A quantum dot connected to only a single lead is often referred to as a single electron box[7]. For such a box to contain a quantized number of electrons the tunnel conductance G of the box to its surroundings must be much less than the conductance quantum $G_Q = 2e^2/h$ (here e is the electron charge and h is Planck's constant)[1]. Charge quantization and thermal fluctuations have been studied experimentally and theoretically in 3-dimensional single electron boxes with fixed tunnel conductance and many weakly transmitting channels.[4-5, 7, 8-11] For 2-dimensional quantum dots in semiconductor systems, the loss of charge quantization due to quantum tunneling has been studied theoretically[12-14] and investigated indirectly in transport measurements on single[15] and multiple[16-20] quantum dots. We report here the direct measurement of the destruction of charge quantization in a 2-dimensional quantum box by tunneling through a single 1-dimensional channel with variable tunnel conductance.

Figure 1 shows a scanning electron microscope photograph (a) and a schematic diagram (b) of the structure, consisting of the electrometer single-electron transistor (SET) on the left, capacitively coupled to the single electron box on the right. Both devices were defined in a GaAs/AlGaAs heterostructure containing a high-mobility ($5\times10^5$ cm$^2$/Vs), near-surface (57 nm) two-dimensional electron gas (2DEG) with sheet density $3.7\times10^{11}$/cm$^2$. Metal gates (the light regions in Fig. 1(a) ) were deposited on the heterostructure surface using electron beam lithography and Cr-Au metallization. When negative voltages are applied to these gates, electrons in the 2DEG layer below are depleted, reversibly defining the contours of the devices in the 2DEG layer. As shown in the schematic, seven independently tunable metal gates were used to define the two devices. Narrow constrictions defined by the central gate $V_c$ and gates $V_{q1}$, $V_{q2}$, $V_{e1}$ and $V_{e2}$ form adjustable quantum point contacts that determine the tunnel-coupling between the electrometer, the box, and the leads. All measurements were done in a He dilution refrigerator cooled to base temperature 25 mK.

Figures 2(a)-2(c) demonstrate the operation of the electrometer as a direct probe of the charge in the single electron box and confirm the advantage of this method over transport measurements. To test the electrometer and determine its charge sensitivity both quantum point contacts to the box were set in the weak tunneling regime, forming a quantum dot, and gate voltage $V_g$ on the dot was varied. Varying $V_g$ would induce a continuously increasing charge on the quantum dot if no tunnel barriers existed, but for weak tunneling, the total number of electrons on the dot must be an integer N. This tension between the induced and actual dot charge produces excess electrostatic energy, resulting in the Coulomb blockade of charge transport through the dot. However, at periodic values of $V_g$ the electrostatic energy of the dot ground state is the same for N or N+1 electrons, permitting N to change and current to flow[1-3], as is evident in the dashed curves for dot conductance $G_{dot}$ in Figs. 2(a)-2(b). At each peak the average charge on the dot changes by one electron; between peaks, the number of electrons on the dot remains constant.

The solid traces in Figs. 2(a)-2(c) show clear, stepwise changes in the electrometer conductance $G_{elect}$ corresponding to single electron changes in the charge on the dot. The vertical jumps in $G_{elect}$ align precisely with the $G_{dot}$ conductance peaks. Figures 2a-2c correspond to progressively lower values of the conductance G through the quantum point contact $V_c$-$V_{q1}$. As G is decreased the electrometer signal $G_{elect}$ remains unchanged, while the transport signal $G_{dot}$ decreases to values below the amplifier noise limit. Thus changes in dot charge in Fig. 2c which would be undetectable in conductance measurements on the dot are clearly evident in the signal from the separate electrometer. The sensitivity of the SET referred to the single electron box can be found for our device by comparing the noise in $G_{elect}$ with the size of single electron steps, giving dq ~ 4 x $10^{-3}$ e/√Hz. SET sensitivities are usually referenced to the charge on the SET gate; for our device, this sensitivity is dq ~ 6 x $10^{-5}$ e/√Hz, which compares well with previous SET devices[1].

Figure 3 shows the direct measurement of the destruction of charge quantization in a single electron box. For these measurements the bottom point contact $V_c$-$V_{q2}$ is completely pinched off by applying a large negative voltage; electrons enter or leave through a channel defined by the upper point contact $V_c$-$V_{q1}$. Under these circumstances no current flows through the box, but charge can again be capacitively induced on the box by applying a voltage $V_L$ to the upper lead. Figures 3a-3f plot the electrometer signal $G_{elect}$ as a function of $V_L$ for values of tunnel conductance in the upper point contact increasing from G = 0.08$G_Q$ in Fig. 3a to G = 1.04$G_Q$ in Fig. 3f. As the tunnel conductance to the box is increased, the single electron steps are smoothed out, and they are completely absent for G = 1.04$G_Q$.

Matveev[4, 12] has considered this situation theoretically by computing the Coulomb interaction energy of a quantum box connected to an electron reservoir by a quantum point contact in the tunneling regime. This energy is shown to evolve continuously from its capacitive value in the weak tunneling limit G << $G_Q$ to precisely zero at G = $G_Q$ where charge quantization and the Coulomb blockade are completely destroyed. In this theory, as G increases from 0 to $G_Q$, the charge steps are rounded and the slopes of the horizontal plateaus increase; this is evident in the data in Fig. 3. Theoretical curves for box charge in the weak[8] and strong[12] tunneling limits are shown as insets for Figs. 3a and 3b and Figs. 3e and 3f respectively. At the center of each horizontal plateau the influence of tunneling is most directly seen, whereas at the vertical sides of the charge steps, where the Coulomb blockade goes to zero, the effect of thermal fluctuations is greatest. Thus we can characterize the strength of quantum fluctuations by the slope $S = dQ_{box}/dQ_{ind}$ of the plateau at its center, where $Q_{box}$ is the actual box charge including the effects of quantization, and $Q_{ind}$ is the charge capacitively induced by $V_L$. For perfect charge quantization $S = 0$; when quantization is completely destroyed, $S = 1$ as for a classical capacitor.

Figure 4 utilizes this quantity S to compare the destruction of charge quantization by quantum mechanical tunneling for a large data set similar to Figs. 3a-3f with theory. In the weak tunneling limit $G << G_Q$, theory predicts that charge is precisely quantized; the charge steps are horizontal and the slope $S = 0$. This is evident in the data in Fig. 4: the slope starts at $S \sim 0$ and rises continuously with G as the charge steps begin to smooth out. In the weak tunneling limit the data rise somewhat more rapidly than perturbation theory calculations[4,8,11]. In the strong tunneling limit $G_Q - G << G_Q$ charge fluctuation theory[12] predicts that charge quantization in a single electron box connected to a reservoir by a one-dimensional channel is completely destroyed at precisely $G = G_Q$. The theoretical slope S vs. G in the strong tunneling limit is shown by the solid curve in Fig. 4 which rises sharply with G then saturates at $G = G_Q$. This predicted behavior is confirmed by our charge measurements: the most dramatic feature of the data in Fig. 4 is a sharp increase of the slope S with tunnel conductance just below $G = G_Q$ followed by an abrupt saturation at $S \sim 1$ for $G > G_Q$. Single electron steps are completely absent in the data above $G = G_Q$, as for Fig. 3f.

The authors thank K.A. Matveev and B.I. Halperin for helpful discussions. This work was supported at Harvard by NSF grant NSF DMR-95-01438, ONR grant N00014-95-1-0866, the MRSEC program of the NSF under award DMR-94-00396, and the DOD NDSEG (D.S.D.) Fellowship program, and at UCSB by grant AFOSR F49620-94-1-0158.

References

[1] H. Grabert & M. H. Devoret, eds, *Single Charge Tunneling*, vol. 294 of *NATO ASI Series B* (Plenum, New York, 1992), and references contained therein.

[2] L. L. Sohn, L. P. Kouwenhoven, and G. Schon, eds, *Mesoscopic Electron Transport*; vol. 345 of *NATO ASI Series E* (Kluwer, Boston, 1997), and references contained therein.

[3] M. A. Kastner, *Rev. Mod. Phys.* **64**, 849-58 (1992), and references contained therein.

[4] K. A. Matveev, *Sov. Phys. JETP* **72** (5), 892-9 (1991).

[5] L. I. Glazman and K. A. Matveev, *Sov. Phys. JETP* **71** (5), 1031-7 (1990).

[6] See for example Chp. 9 of ref. 1.

[7] P. Lafarge, H. Pothier, E. R. Williams, D. Esteve, C. Urbina, and M. H. Devoret, *Zeit Phys. B* **86**, 327-32 (1991).

[8] H. Grabert, *Phys. Rev. B* **50**, 17 364-77 (1994).

[9] X. Wang, *Phys. Rev. B* **55**, 4073-6 (1997).

[10] G. Falci, J. Heins, G. Schon, G. T. Zimanyi, Physica B **203**, 409-16 (1994).

[11] W. Hofstetter and W. Zwerger, *Phys. Rev. Lett.* **78**, 3737-40 (1997).

[12] K. A. Matveev, *Phys. Rev. B* **51**, 1743-51 (1995).

[13] K. A. Matveev, L. I. Glazman, and H. U. Baranger, *Phys. Rev. B.* **53**, 1034-7 (1996).

[14] J. M. Golden and B. I. Halperin, *Phys. Rev. B* **53**, 3893-900 (1996).

[15] L. P. Kouwenhoven, N. C. van der Vaart, A. T. Johnson, W. Kool, C.J.P.M. Harmans, J.G. Williamson, A.A.M. Staring, and C.T. Foxon, *Z. Phys. B* **85**, 367-73 (1991).

[16] F. R. Waugh, M.J. Berry, D.J. Mar, R.M. Westervelt, K.L. Campman, and A.C. Gossard, *Phys. Rev. Lett.* **75**, 705-8 (1995).

[17] L. W. Molenkamp, K. Flensberg, and M. Kemerink, *Phys. Rev. Lett.*, **75**, 4282-5 (1995).

[18] C. Livermore, C. H. Crouch, R. M. Westervelt, K. L. Campman, and A. C. Gossard, *Science*, **274**, 1332-5 (1996);

[19] A. S. Adourian, C. Livermore, R. M. Westervelt, K. L. Campman, and A. C. Gossard, *Superlattices and Microstructures* **20**, 411-17 (1996).

[20] C. H. Crouch, C. Livermore, R. M. Westervelt, K. L. Campman, and A. C. Gossard, *Appl. Phys. Lett.* **71** (6), 817-19 (1997).

Figure Captions

FIG. 1. (a) Scanning electron microscope photograph of electrometer and single electron box. Bright areas are metal surface gates; dark area is the GaAs heterostructure surface. Rectangular box encloses gates used for this experiment. Scale bar is 1μm. (b) Wiring schematic of the device, consisting of four quantum point contacts (defined by the central gate $V_c$ and gates $V_{q1}$, $V_{q2}$, $V_{e1}$ and $V_{e2}$) and two confining side walls which act as capacitively coupled gates (labeled $V_e$ and $V_g$), also independently controllable. When these gates are energized, the electrometer and box are formed in the two-dimensional electron gas (2DEG) 57 nm beneath the surface.

FIG. 2. Differential conductance $G_{elect}$ (solid curves) of electrometer and $G_{dot}$ (dashed curves) of the dot measured simultaneously vs. gate voltage $V_g$. The electrometer is biased on the side of a Coulomb blockade conductance peak and a cancellation signal is added to $V_e$ in order to compensate for the capacitive coupling between the electrometer and gate $V_g$, so that the electrometer conductance $G_{elect}$ is sensitive only to the charge on the dot. The cancellation is adjusted to give horizontal plateau centers at $G \sim 0$. The number of electrons on the dot changes by one at each peak in $G_{dot}$ causing $G_{elect}$ to rise in steps. Plateaus in $G_{elect}$ between peaks indicate

the dot charge is constant in these regions. In (a) the conductance G of quantum point contact $V_c$-$V_{q1}$ is most open ($G = 0.14G_Q$, where $G_Q = 2e^2/h$) resulting in relatively high conductance peaks for the dot. In (b) the point contact is slightly more pinched off ($G = 0.04G_Q$), and In (c) the point contact is completely pinched off with $G \sim 0$; here conductance peaks are no longer perceptible. In all three cases the electrometer signal is equally strong indicating that the electrometer is a more sensitive probe of the dot charge than transport measurements.

FIG. 3. Electrometer conductance $G_{elect}$ vs. voltage $V_L$ applied to the upper lead of the box. (a) The single electron steps are nearly horizontal for a weakly tunnel-coupled box with $G = 0.08G_Q$; the steps are progressively smeared in (b) to (e) at conductance values (b) $G = 0.27G_Q$; (c) $G = 0.45G_Q$; (d) $G = 0.66G_Q$; (e) $G = 0.85G_Q$; in (f) the steps are entirely absent at a conductance value of $G = 1.04G_Q$ (where $G_Q = 2e^2/h$). Insets are theoretical calculations for single steps: in (a) for $G = 0$ and (b) for $G = .25G_Q$, at $T \sim 100mK$[8]; in (e) for $G = .85G_Q$ and (f) for $G = G_Q$ at $T = 0K$[12].

FIG. 4. Average slope $S = dQ_{box}/dQ_{ind}$ ($Q_{box}$ and $Q_{ind}$ are the actual and induced box charge respectively) of single electron steps at the midpoint of the plateaus vs. tunnel-conductance G of the box to the lead. Each data point in Fig. 4 is derived from the average of many sweeps of $G_{elect}$ vs. lead voltage $V_L$. The slope S was found using the vertical spacing of single electron steps to calibrate changes in $Q_{box}$ from $G_{elect}$ and the horizontal step spacing to calibrate changes in $Q_{ind}$ from $V_L$. The conductance G of the upper box point contact was calibrated vs. gate voltage $V_{q1}$ by measuring G vs. $V_{q1}$ with the lower point contact open for several values of $V_{q2}$, followed by an extrapolation to the case where the lower point contact is completely pinched off.

The error bars are estimates of the total statistical and systematic error. Solid gray curves are the theoretical predictions in the weak tunneling limit[4] (3-dimensional theory is applicable in this limit) and in the strong tunneling limit[12] for values of G near $G_Q$. The slope $S \sim 0$ for $G \ll G_Q$ and rises sharply just below $G = G_Q$, saturating at $S = 1$ for $G > G_Q$ as predicted by theory.

**(a)**

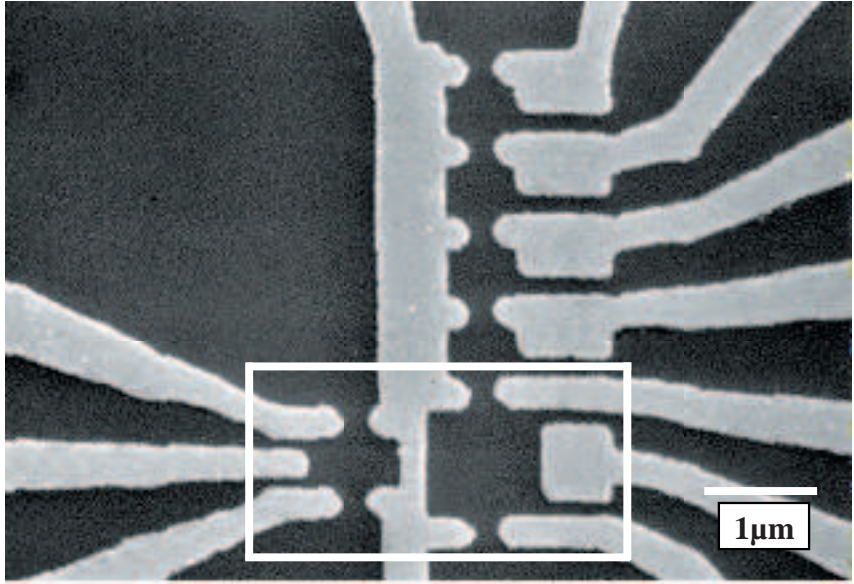

**(b)**

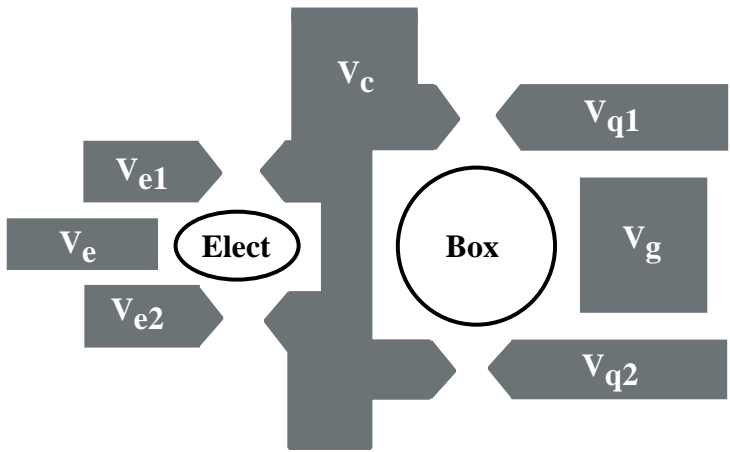

D. S. Duncan *et al.*
Figure 1 of 4

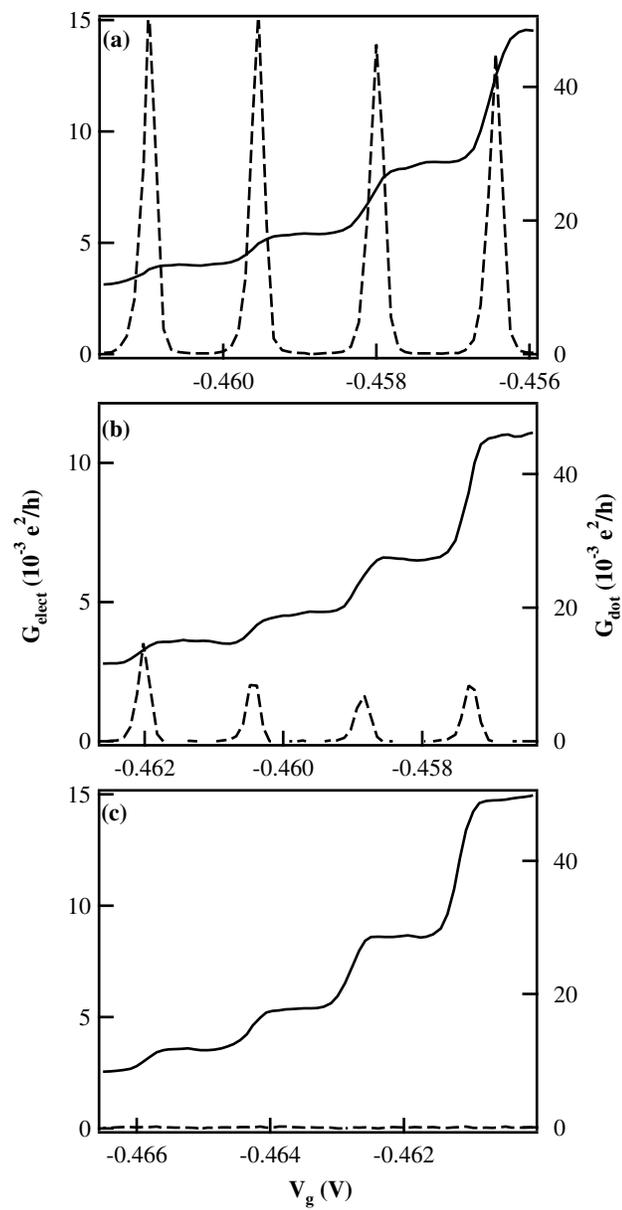

D. S. Duncan *et al.*
Figure 2 of 4

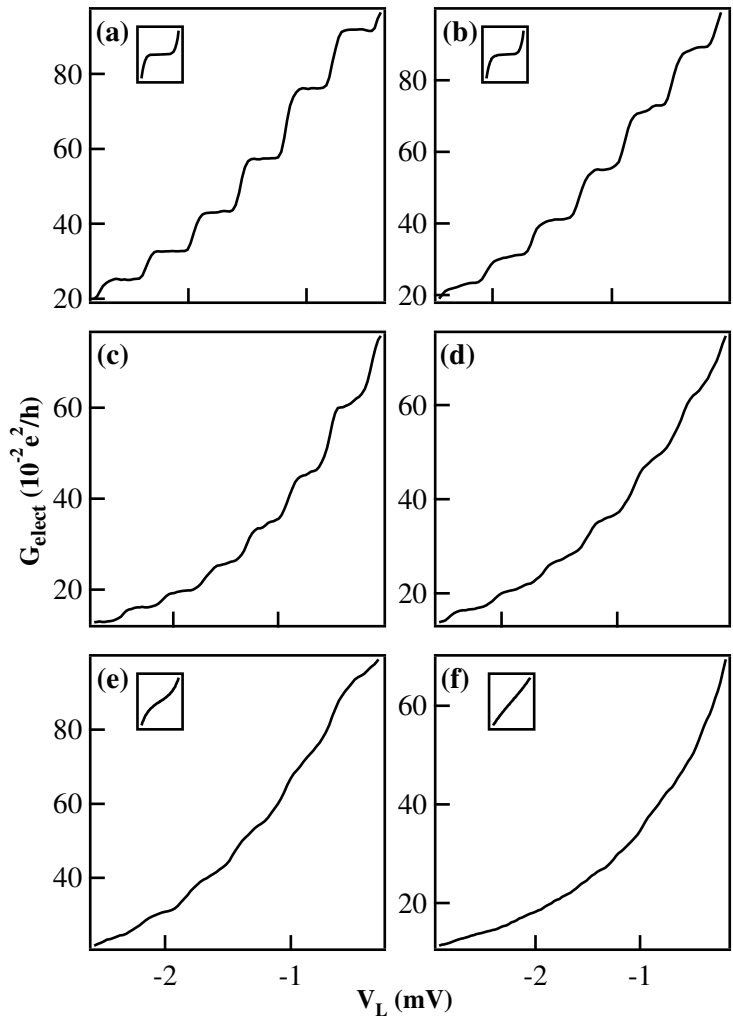

D. S. Duncan *et al.*
Figure 3 of 4

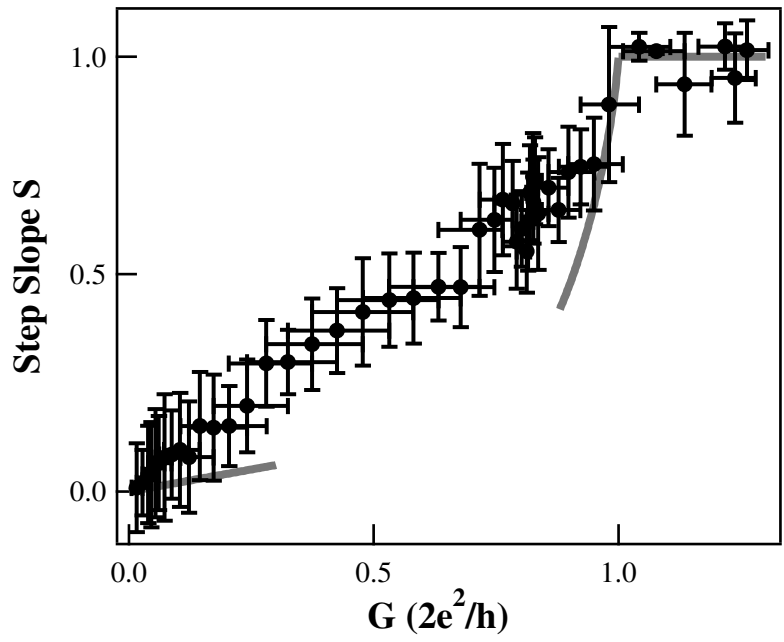

D. S. Duncan *et al.*
Figure 4 of 4